\documentclass[%
mathleft,%
final,%
]{an}
\usepackage{graphicx}

\usepackage{times}
\newcommand{\scrbox}[1]{\ensuremath{{\mbox{\scriptsize #1}}}}

\newcommand{\teff}{{\ensuremath{T_{\scrbox{eff}}}}}
\newcommand{\Msol}{\ensuremath{\,\mbox{\it M}_{\odot}}}

\newcommand{\Dturb}{\ensuremath{D_{\scrbox{T}}}}
\begin{document}

\Pagespan{1}{}
\Yearpublication{2013}%
\Yearsubmission{2012}%
\Month{1}%
\Volume{334}%
\Issue{1}%
\DOI{This.is/not.aDOI}%

\title{The stratified evolution of a cool star}

 \author{G. Michaud \inst{1,2}\fnmsep\thanks{Corresponding author. 
  \email{michaudg@astro.umontreal.ca}} 
	\and J. Richer
          \inst{2}
		\and O. Richard
					\inst{3}
          }
	\titlerunning{Stratified evolution}
\authorrunning{G. Michaud \& J. Richer \& O. Richard}

   \institute{LUTH, Observatoire de Paris, CNRS, Universit\'e Paris Diderot,
     5 Place Jules Janssen,
     92190 Meudon, FRANCE
     \and  
   D\'epartement de Physique, Universit\'e de Montr\'eal,
       Montr\'eal, PQ, H3C~3J7, CANADA\
			\and
Universit\'e Montpellier II - GRAAL, CNRS - UMR 5024, place Eug\`ene Bataillon, 
34095 Montpellier, FRANCE\\}

\received{\today}
\accepted{XXXX}
\publonline{XXXX}

\keywords{stars: interiors -- stars: evolution -- stars: abundances -- stars: horizontal--branch}

\abstract{%
A low mass star usually experiences stratification and abundance anomalies during its evolution. A 0.95 \Msol{} star with a metallicity $Z = 0.004$ is followed from the main--sequence to the Horizontal Branch (HB).  On the main--sequence the larger effects of stratification may come from accretion as was  suggested in relation to metallicity and  planet formation.  As it evolves through the giant branch, stratification appears around the hydrogen burning shell. It may create hydrodynamic instabilities and be related to abundance anomalies on the giant branch.  After the He flash the star evolves to the HB.  If it loses enough mass, it ends up a hot HB star (or in the field an sdB star) with effective temperatures larger than 11000\,K. All sdB stars are observed to have an approximately  solar iron abundance   whatever their original metallicity, implying overabundances by factors of up to 100.  So should the 0.95 solar mass star.  How its internal hydrodynamic properties on the main sequence may influence its fate on the HB is currently uncertain. }

\maketitle

\section{Astrophysical context}
In most stellar mass intervals, stars are significantly stratified by particle transport processes   during part of their evolution.  This is in addition to stratification due to nuclear burning.  For instance, as a 0.95\,\Msol{} star of metallicity $Z= 0.004$ remains
 cooler than 6500\,K on the main--sequence, the larger effects of potential stratification come possibly from accretion but are limited by thermohaline mixing.  This could be linked to planet accretion and the destruction of Lithium.

As it ascends the giant branch, effects are very small at the surface but might be larger close to the H burning shell.  
The largest stratification occurs on the horizontal branch (HB) but is strongly dependent on the effective temperature that the star ends up with.  The spread of HB masses that leads to the effective temperature spread on the HB is potentially linked to the low mass star structure on the main--sequence.

We here review the stratification during the evolution of such a star using mainly results obtained with an evolutionary code that takes into account transport by atomic diffusion outside of convection zones.  Calculations  were carried out with all usual equations of standard stellar evolution with in addition a set of 56 coupled differential equations to take into  full acount the effects of the atomic diffuson of the 28 species included.  Radiative acceleration and Rosseland averaged opacity are continuously recalculated during evolution so that the calculations are continuously consistent with all abundance variations as described in Richer et al. (\cite{RicherMiRoetal98}) and Turcotte et al. (\cite{TurcotteRiMietal98}) where the only parameter of the calculations, the mixing length, is fixed by properties of the Sun.  Parameters are introduced only to describe macroscopic physical processes competing with atomic diffusion.

The structure of this review follows the major evolutionary stages of a low mass star;
we are not concerned with stratification originating directly from nuclear reactions.  It is stratification caused by transport processes that is looked into here.

\section{Main sequence evolution of a cool star}
\label{sec:MainSequence}
A 0.95\Msol{} star of original metallicity $Z= 0.004$ was followed from the pre$-$main$-$sequence to the He flash.  As may be seen from Fig.\, 1 of Michaud, Richer \& Richard (\cite{MichaudRiRi2010}) the effects of diffusion on the HR diagram become apparent mainly around turnoff as well as  close  to the hook, where on the giant branch the H burning shell crosses the point of deepest inward expansion of the  surface convection zone.  These effects are small.  

\begin{figure}
\includegraphics[angle=0, width=1.0\linewidth]{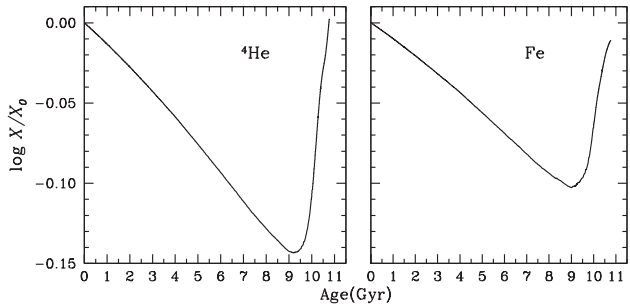}
\caption{Surface abundances of He and Fe during the evolution of a 0.95\Msol{} star from the ZAMS to the He flash.  The settling is progressive during evolution and the underabundance is by the largest factor (-0.15 dex for He) at turnoff.  The He abundance returns to its original value as the He flash is approached but the Fe abundance remains -0.01 dex below as some Fe is concentrated in the stellar center.}
\label{fig:surface}
\end{figure}

\begin{figure}
\includegraphics[angle=0, width=1.0\linewidth]{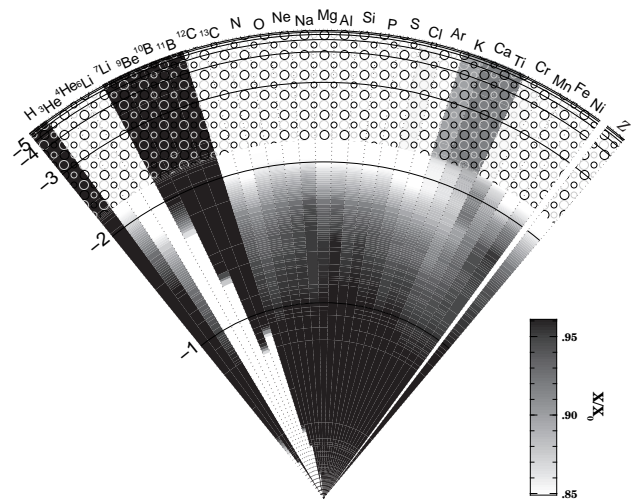}
\caption{The interior concetrations of the 28 calculated species are shown intensity coded  at an age of 8.6\,Gyr.  The radial coordinate is the radius and its scale  is linear, but the logarithmic value of the mass coordinate above a number of points, $\log \Delta M/M_{*}$, is shown on the left of the 
	horizontal black line.  The concentration scale is given in the right insert.  
	  Small  circles  mark the surface  convection zone. The effect of radiative accelerations is evident from K to Mn slightly below the convection zone. It is small as will be the case throughout the main$-$sequence. }
\label{fig:interior}
\end{figure}

The  variations of the surface abundances of He and Fe are shown in Fig.\ref{fig:surface}.  They were calculated for all 28 species but  the reduction factors are similar for most metals.  The depth of the convection zone causes the radiative accelerations to play a small role which may  be seen from Fig.\,\ref{fig:interior} where the interior concentrations for all 28 calculated species are shown at an age of 8.6\,Gyr.
The interior concentration variations caused by diffusion processes are by around 0.1\,dex (note that the concentration scale on the right inset of the figure ranges from a factor of 0.85 to 0.95).  For light elements (from H to O) nuclear raections are the main cause of concentration changes and are concentrated in the inner 50\,\% by radius.  In the outer 50\,\% of the radius the changes are due  to atomic diffusion which is here dominated by gravitational settling.  However one sees the effect of radiative accelerations below the convection zone from K to Mn.

\subsection{Metallicity around 13\,Gyr}
\label{sec:Metallicity13}
The Pop II  stars with the lowest metallicity (see for instance Beers \& Christlieb \cite{BeersChristlieb2005}) are thought to have formed some 13\,Gyr ago, a few hundred Myr after the Big Bang.  
\begin{figure}
\centerline{\includegraphics[angle=0,width=0.8\linewidth]{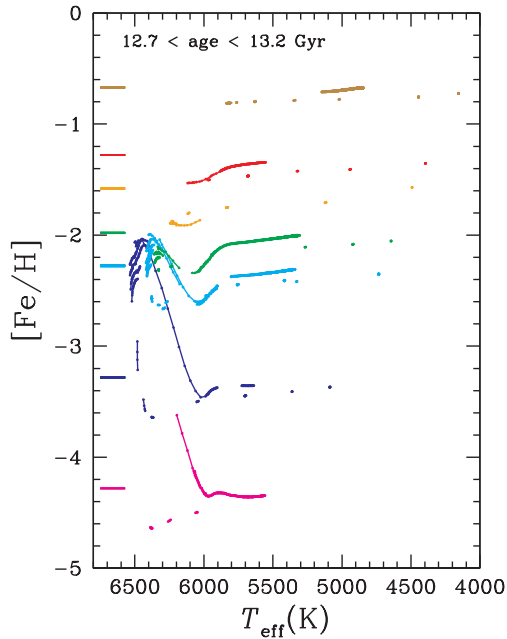}}
\caption{The colored line segments represent the iron abundance between 12.6 and 13.2\,Gyr in stars of different mass and different original metallicity.  Stars which are before turnoff appear as a point, since slow evolution causes the segment to collapse to a point.  However those which are in more rapid evolutionary stages (around turnoff or on the subgiant branch) appear as line segments.  Note that subgiant branch stars have slightly larger abundances than main--sequence stars of the same \teff. The initial Fe abundance is indicated by a small line on the left. }
\label{fig:metallicity}
\end{figure}
They may be used as messengers from the earliest times.  But to what extent are we certain that their surface composition we see today is that with which they formed.  The 0.95 \Msol{} star we follow will not see its surface abundances modified strongly by atomic diffusion.  But this is not generally the case for halo fields stars as may be seen in Fig.\,\ref{fig:metallicity}.  As one considers stars with the lowest metallicities, some of the surviving ones may have effective temperatures larger than 5800\,K.  If there is no process outside of convection zones competing wih atomic diffusion, surface overabundances of iron by more than a factor of ten are possible so that the apparent metallicity would be larger than the one the star formed with.  This could be detected since iron to carbon or to oxygen ratios would appear much larger than solar since the carbon and oxygen abundances are not expected to be larger than the original one (see Fig.\,9, 10 and 11 of Richard, Michaud \& Richer \cite{RichardMiRi2002}).  Most but not all observed very low metallicity stars have an effective temperature smaller than 5800\,K.

\subsection{Accretion of planets}
\label{sec:Accretion}
There is now ample evidence for the existence of planets around most stars.  Through disk migration and tidal interactions, 
planets move inwards leading to accretion (e. g. Jackson, Barnes \& Greenberg \cite{JacksonBaGr2009}). Many of the observed innermost exoplanets are expected to be accreted within a few Gyrs.  Since planets have larger metallicity than the stellar cloud from which they formed, they should increase the metallicity  of the accreting star.  On the other hand stars that have been observed to have planets have been found to have larger metallicity than those without planets (Fischer \& Valenti \cite{FischerVa2005}).  
Is this overmetallicity of stars with planets due to the accretion of planets or is an overmetallicity of the  original nebula from which the star and planet formed the cause of planet formation?  In other words, is the overmetallicity primordial or due to planet accretion?  The answer to this question depends on the mass in which the planet mixes after merging with the star.

Vauclair (\cite{Vauclair2004}) suggested that the larger metallicity of the matter accreted from the planet implied that, due to thermohaline mixing, the  mass into which the planet mixed was generally much larger than the mass of the convection zone.  Consider a planet with a 1.0 Jupiter mass accreting to a 1.4\Msol{} star. Garaud (\cite{Garaud2011})  found that (see her Fig.\,2) within $10^4$\,yr, the depth of the convection zone adjusts so that the Ledoux criterion is satisfied at its bottom; overmetallicity has been reduced by 20\,\%.  Thermohaline convection  further, but more slowly, extends the mixing.  After $10^7$\,yr, overmetallicity has been reduced by a factor of ten so that the effect of the accretion has virtually disappeared from the surface.  
Consequently, the increased metallicity observed on planet bearing stars has to be of primordial origin.  It cannot have been caused by  planet accretion since the effect of the accretion lasts for only a small fraction of the star's evolution time.  Planet accretion cannot be an important cause of star stratification (see also Proffitt \& Michaud \cite{ProffittMi89a}). These results depend, however, on an evaluation of the efficiency of thermohaline convection.

\subsection{Thermohaline convection}
\label{sec:Thermohaline}
Thermohaline convection is for instance observed in oceanic  waters when there is salt water above cooler fresh water.  There it is often called \textit{salt fingering}. In stars the increased metallicity plays the role of the increased salinity.  In the absence of a proper calculation of the transport efficiency of thermohaline convection in stars, one has often used a mixing length evaluation.  It is however very uncertain since it depends rather sensitively on the assumed elongation of the cells.  Different evaluations differed by a factor of $\sim$\,100. More recently  detailed simulations first in 2-D (Denissenkov \cite{Denissenkov2010}) and then in 3$-$D (Traxler, Garaud \& Stellmach \cite{TraxlerGaSt2011}) have improved the situation.  The results of the two simulations are in agreement.

In their 3$-$D simulation, Traxler, Garaud \& Stellmach (\cite{TraxlerGaSt2011})  determine the turbulent diffusivity in a box where they impose a stably stratified thermal field but an unstably startified compositional field.  They carried out simulations at resolutions of $96 \times 96 \times 192$ to $384 \times 384 \times 384$ in order to approach as much as possible the parameters appropriate for the stellar case.  While they were still far from it, they claim to have arrived at an asymptotic value for the turbulent diffusivity that may be used as a universal law.  As shown in their Fig.\,3, they agree with the evaluation of Kippenhahn, Ruschenplatt and Thomas (\cite{KippenhahnRuTh80}) which is the lower of the evaluations previously considered.
While their results seem to converge to an asymptote as the ratio of thermal to compositional diffusivity and the Prandtl number approach stellar values, they are still far from them and higher resolution simulations would still appear needed to firmly establish the results. Their results were used in the discussion of \S\,\ref{sec:Accretion} and have some impact also on the transport above the H burning shell of giants described in the next section as well as for the Li abundance in stars (Theado \& Vauclair \cite{TheadoVa2012}).

\section{Giant Branch}
\label{sec:Giant}
\begin{figure}
\includegraphics[angle=0, width=0.9\linewidth]{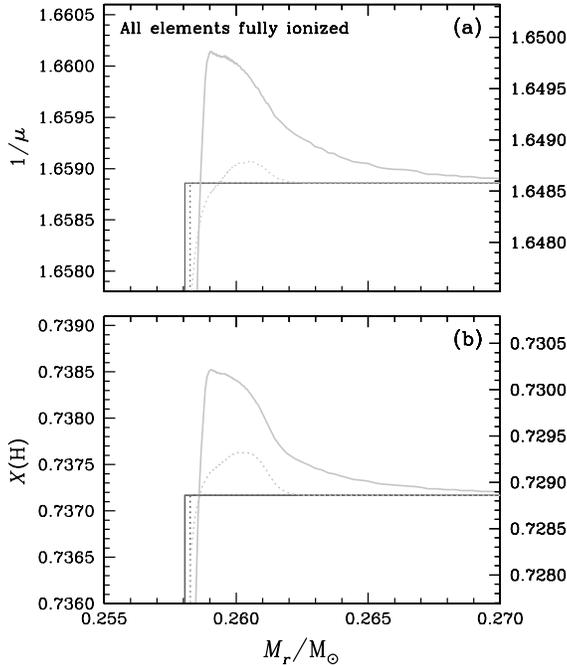}
\caption{
\emph{Upper panel} Interior profile of $\mu$ as a function of the interior mass, $M_r$, immediately after the first dredge--up (black lines) and (gray lines)  as the hydrogen  burning shell crosses  the region of the composition break left by the first dredge--up (more precisely $\sim 1.5$\,Myr before). The solid lines represent the calculations with atomic diffusion and the dotted lines those without.  The solid lines refer to the scale on the left and the dotted lines to the scale on the right (the two scales were adjusted so that the two horizontal black lines are superposed).  The two scales cover the same $\Delta (1/\mu)$ interval.  \emph{Lower panel} Interior profiles of the corresponding $X(\rm{H})$ values.   }
\label{fig:giants}
\end{figure}

As the 0.95\Msol{}, $Z= 0.004$ star goes up the giant branch, its hydrogen burning shell approaches the depth where the surface convection zone had its deepest extension (at the first dredge--up).  Concentrations were homogenized from the surface down to that depth but there appeared a discontinuity in many concentrations at that depth. As the hydrogen burning shell crosses that discontinuity, the star describes what is often called the \textit{hook} in the HR diagram.  
 As the hook is approached there occurs, slightly ahead of the H burning shell, an H abundance increase   caused  by $^3\rm{He}$ burning (the dotted gray line on the lower panel of Fig.\,\ref{fig:giants}).  This  H abundance translates into a $\mu$ gradient inversion, shown as the dotted gray line in the upper panel of Fig.\,\ref{fig:giants}, which has been suggested to cause mixing, by thermohaline convection,  between the burning shell and the surface (Eggleton, Dearborn \& Lattanzio \cite{EggletonDeLa2006}; Charbonnel \&  Zahn \cite{CharbonnelZa2007}).  This could explain abundance variations seen on the giant branch.

  However atomic diffusion starts to affect the homogenized concentrations  as soon as the convection zone starts receding.  Does atomic diffusion have a significant effect on the $\mu$ gradient inversion that drives the thermohaline convection?

On the lower panel of Fig.\,\ref{fig:giants} are shown (from data used by Michaud, Richer \& Richard  (\cite{MichaudRiRi2010}) where more details may be found; see in particular their Fig.\,8) the $X(\rm{H})$ interior profiles immediately after the convection zone starts receding (black lines) and just before the H burning shell crosses the region of the compositional break (gray lines). Solid lines represent calculations with diffusion and dotted lines those without.  The H abundance change   is $\sim$\,3 times larger in the model with diffusion than in that without diffusion while the effect on $\mu$ is by a factor of $\sim$\,6. It would then appear that the gravitational settling of He leading to the H abundance increase has a larger effect than  $^3$He burning.

The  turbulence induced by thermohaline convection tends to counteract gravitational settling. 
 The concentration gradients of metals and of He (or H) implied by the values of $X(\rm{H})$ on Fig.\,\ref{fig:giants} are very small, yet it is only through those gradients that turbulence ($\Dturb$) has an effect, whereas atomic diffusion also acts through the much larger $g$ driving terms.  Consequently, $\Dturb$ can have an effect only if it is much larger than the atomic diffusion coefficient, $D_{ip}$.  But how much larger?  To have a significant effect on the diffusion velocity, $\Dturb$ must lead to a contribution similar to the driving terms of atomic diffusion.
 Turbulence then has an effect if
\begin{equation}
	\Dturb\left|  \frac{\partial \ln X_{i}}{\partial r}\right| \sim D_{ip} \left| \frac{A_i m_p}{kT}g\right|.
	\label{eq:turbv}
\end{equation}
Eq.\,(\ref{eq:turbv}) was evaluated roughly using drift velocities of He  and the value of $X(\rm{He})$ from Michaud et al (\cite{MichaudRiRi2010}).  The  \Dturb{} so obtained from Eq.\,(\ref{eq:turbv}) is shown on Fig.\,9 of that paper.  It varies 
between $10^6$, immediately above the burning shell, and $10^7$\,cm$^2$/s immediately below the surface convection zone.  According to Sect.\,\ref{sec:Thermohaline} this upper limit is two orders of magnitude larger than the best evaluations of the thermohaline convection turbulent diffusion coefficient.  The gradients that atomic diffusion leads to should then easily be maintained in the presence of this turbulence.

Whether the $\mu$ inversion has any importance for surface abundances also depends on the mixing coefficient that thermohaline convection leads to.  The lower value of the mixing coefficient  discussed in Sect.\,\ref{sec:Thermohaline} is below what is required to cause sufficient mixing with the surface according to Garaud  (\cite{Garaud2011}).  That conclusion is however based only on  the contribution  of $^3$He burning to the $\mu$ gradient inversion since atomic diffusion was not included in the calculations she refers to.  Would the larger $\mu$ gradient values that atomic diffusion leads to make a difference?  A calculation including atomic diffusion and thermohaline convection with the coefficient  described in Sect.\,\ref{sec:Thermohaline} is needed to settle this question.

\section{Horizontal branch and sdB stars}
\label{sec:Horizontal}
As a 0.95\Msol{} star leaves the giant branch, it proceeds to the horizontal branch (HB).  Some such stars loose more  mass than others through a process which is not currently well understood.  Depending on the mass it looses the star ends up with a different \teff{} on the HB.  In Pop II stars, some HBs of globular clusters are populated to relatively high \teff{} and the field equivalent, the sdB stars, extend into the sdO stars. 

 \begin{figure}
\centerline{%
\includegraphics[angle=0, width=0.8\linewidth]{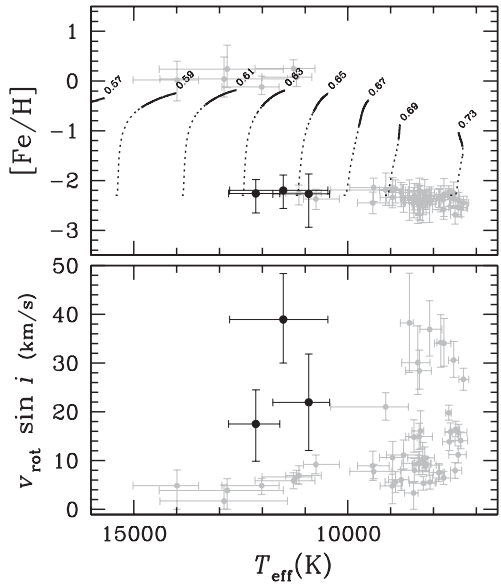}}
\caption{\emph{Upper panel} Observed Fe abundances in HB  stars of M15, M68 and  M92. The black line segments represent the iron abundances during the HB.  Dotted part, first 10\,Myr and solid part from 10 to 30 Myr. \emph{Lower panel} Corresponding rotational velocities. Gray points are data from Behr (\cite{Behr2003}) while the three black ones  identify rapidly rotating ones. }
\label{fig:Fe_HB}
\end{figure}

In globular clusters, HB stars should have the same composition as on the giant branch of the cluster except perhaps for species involved in the CNO cycle and affected by proton burning.  The abundance of most metals, and in particular of iron, should be the same as on giants.  However it had been predicted that atomic diffusion should cause abundance anomalies  on the HB (Michaud, Vauclair \& Vauclair \cite{MichaudVaVa83}).  This prediction has now been strikingly confirmed in all globular clusters with a HB extending to $\teff > 11000$\,K.  As shown on Fig.\,\ref{fig:Fe_HB} for the clusters M15, M68 and M92, HB stars hotter than 11000\,K have approximately a solar iron abundance while the cooler ones have the same Fe abundance as giant stars of the cluster ($\sim 1/200$ solar).  Similarly in clusters with different metallicities (e.g. M3, M13, NGC288, NGC6752, NGC1904, NGC2808), HB stars with $\teff > 11000$\,K have about a solar iron abundance while the lower \teff{} stars have the same iron abundance as giants (Behr \cite{Behr2003}, Moehler et al \cite{MoehlerSwLaetal2000}, Fabbian et al \cite{FabbianReGretal2005}, Pace et al \cite{PaceRePietal2006}).  The  black points on Fig.\,\ref{fig:Fe_HB} show a few exceptions:  a few more rapidly rotating stars around 11000\,K.  The black line segments show the calculated surface abundances of iron in 8 HB models covering the \teff{} range.  The dotted part covers the first 10 Myr of HB evolution and the solid part, the following 20 Myr.  The observed Fe abundances are as expected for most stars with $\teff > \,11000$\,K.  For the cooler stars and for the three represented by black points around 11000\,K, rotation (lower panel of Fig.\,\ref{fig:Fe_HB}) appears to compete succesfully with the diffusion driven by radiative forces.  It has been suggested that, given the observed $v \sin i$ on the HB, meridional circulation is efficient in reducing abundance anomalies in stars cooler than 10000\,K but not in those hotter than 11000\,K with a buffer zone in between (Quievy et al \cite{QuievyChMietal2009}).

\begin{figure}
\includegraphics[angle=0, width=1.0\linewidth]{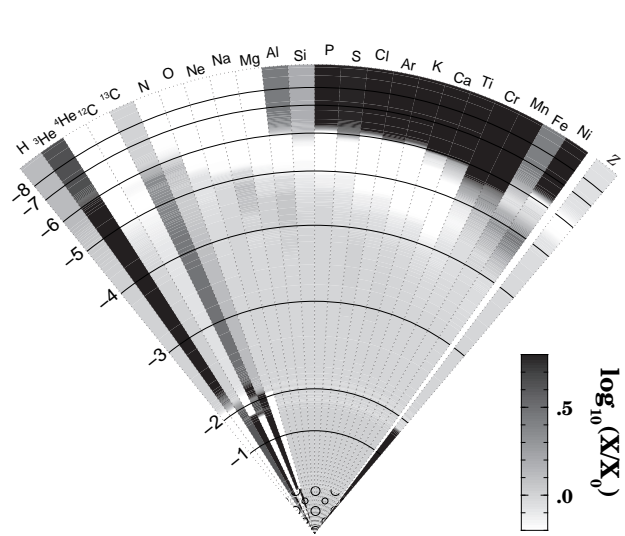}
\caption{The interior abundances of the 23  species included for the HB calculations are shown intensity coded  at an age of 31.1\,Myr after the beginning of the HB in a $\sim 25000$\,K HB star.  The radial coordinate is the radius and its scale  is linear;  the logarithmic value of the mass coordinate above a number of points, $\log \Delta M/M_{*}$, is shown on the left of the 
	horizontal black line.  The concentration scale is given in the right insert.  
	  Small  circles  mark the central convection zone. The effect of radiative accelerations is evident from Ne to Ni. Note that the concentration scale is different from that used for the  main$-$sequence star (see Fig.\,\ref{fig:interior}) .}
\label{fig:interiorHB}
\end{figure}

The calculations used in Fig.\,\ref{fig:Fe_HB} are the continuation of those described in the preceding sections.  In pursuing to the HB, approximately the same procedure as used by  Sweigart\,(\cite{Sweigart87}) was followed. After the He flash, this procedure involves removing a varying amount of mass in order to arrive at different \teff{}s on the HB.   
When on the HB the model is reconverged and, in the models described here, atomic diffusion is allowed to proceed.  In these calculations, the outer $10^{-7}\Msol$ was assumed to be mixed by some turbulent process.  This is the one adjusted parameter for these calculations and it has the same value for all HB and sdB stars discussed here.  An example of the internal abundance distribution this leads to is shown in Fig.\,\ref{fig:interiorHB} (from Michaud, Richer \& Richard \cite{MichaudRiRi2011}).  The effects of atomic diffusion are not only superficial but are felt over the outer third of the radius of the star.  The abundance variations are large, by factors of order ten, and one sees clearly the effect of closed electronic shells on radiative accelerations in the shifting inwards of the local maxima of concentrations from Ca to Ni.

\begin{figure}
\includegraphics[angle=0, width=0.9\linewidth]{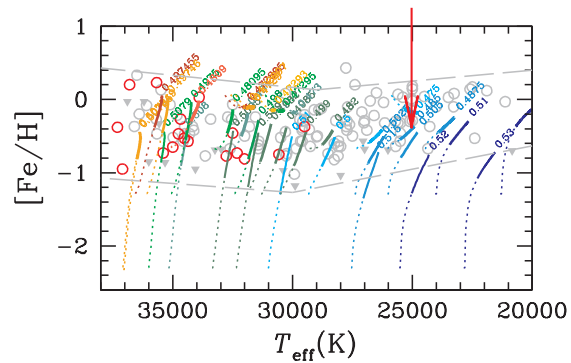}
\caption{Observed Fe abundances in sdB stars (circles) and upper limits (downward pointing triangles) from  Geier  et al (\cite{GeierHeEdetal2010}) who observed 139 sdBs with high resolution spectrographs.  All stars were verified to be slowly rotating.The line segments are models calculated with original Fe abundances of [Fe/H] = -2.3, -1.3, -0.4 and 0.0.  The dotted part of the segment is from 0 to 10 Myr after the beginning of the HB while the solid segment is from 10 to 32 Myr.  The red arrow shows the position of a star originally of 0.95\,\Msol{} with a metallicity $Z = 0.004$.  The masses on the HB are in small characters but one may zoom on them in the online version.  }
\label{fig:Fe_sdB}
\end{figure}
On Fig.\,\ref{fig:Fe_sdB} the calculated Fe surface abundaces are compared to the observed values in sdB stars.  Some 60 evolutionary HB models were calculated with metallicities of $Z_0 = 0.0001$, 0.001, 0.004 and 0.02.  These cover the metallicity range with which one may expect most field sdB stars to have formed.  During their evolution, even most of those starting with a metallicity 200 times smaller than solar end up, after 10 Myr, within a factor of ten of solar   (the solid part of the line segments). This is also the interval were most observed values are found.  The evolutionary calculations were stopped after some 32 Myr on the HB for technical reasons.  The observed sdB stars beyond that time  were identified  in Fig.\,2 of Michaud et al (\cite{MichaudRiRi2011}) by red circles and these are also used to identify those objects in the (Fe, \teff) plane.  The observed upper limits (the inverted gray triangles) would correspond to the first 10 Myr of the HB evolution.

Similar observations were carried out by the same observers for He, C, N, O, Ne, Mg, Al, Si, P, S, Ar, K, Ca, Ti and Cr  and compared with model calculations with a similar level of agreement as for Fe (see Figs. 15 and 16 of Michaud et al \cite{MichaudRiRi2011}).

\section{Conclusion: what do we learn from HB stars?}
\label{sec:Conclusion}
As seen in the preceding Section, complete evolutionary models taking atomic diffusion  and radiative accelerations into account predict approximately the observed metal abundances on hot HB and on sdB stars.  Only one parameter was adjusted and the same value of the parameter was used for all calculations.  This confirms that radiative accelerations are the main cause of the abundance anomalies observed on these objects.  What determines the size of the anomalies is the mass between the surface and the layer where the separation takes place.  This mass could be determined by some turbulent process and lead to the type of solution used here.  But when an adjustable parameter is involved, in this case the mixed mass, there always remains the possibility that a different process would have the same effect. Indeed, it was found for Pop I stars (Vick et al \cite{VickMiRietal2010}, Michaud, Richer \& Vick \cite{MichaudRiVi2011}) that similar surface abundance anomalies are caused at the surface if the process competing with atomic diffusion is either mass loss or turbulence so long as the mass loss rate is such that most of the chemical separation occurs at  the same mass below the surface as is mixed in the turbulent model.   Can asteroseismology  distinguish between the two possibilities?  

Based on a parameter free equilibrium model, pulsations were predicted to occur in some sdB stars 
(Charpinet et al \cite{CharpinetFoBretal96}; Fontaine et al \cite{FontaineBrChetal2003}).  They are caused by the Fe accumulation
where it is the main contributor to opacity. The pulsations were observed shortly after being predicted

The seismic properties of sdB stars were calculated for both models with mass loss and models with turbulence as the competing process by Hu et al (\cite{HuToGletal2011}).  They conclude that the models with turbulence, such as those discussed in the preceding sections, ought to be preferred.  A more detailed fit of the pulsating properties of individual stars than they attempted would seem to be 
needed before this conclusion is definitively accepted. 

Two aspects of hot HB and sdB stars remain puzzling:  what determines the varying amount of mass lost from the giant branch to the HB and why do only the hot stars rotate slowly.  Both of these properties might well be linked to properties of the stars while on the lower main sequence and possibly to the presence of a magnetic field for instance (Charbonneau \cite{Charbonneau2005}).  However, onee the existence of slowly rotating sdB and hot HB stars is accepted,  their abundance anomalies are expected.  
\acknowledgements
   This research was partially supported at  the Universit\'e de Montr\'eal 
by NSERC. We thank the R\'eseau qu\'eb\'ecois de calcul de haute
performance (RQCHP)
for providing us with the computational resources required for this
work.

%

%







\end{document}